\documentclass[apj, revtex4]{emulateapj}
\slugcomment{Accepted to ApJ: February 24, 2014}
\usepackage{float}
\usepackage {microtype}
\usepackage{subfigure}

\begin{document}
\title{Investigating Active Galactic Nuclei Variability Using Combined Multi-Quarter Kepler Data}
\author{Mitchell Revalski \altaffilmark{1}, Dawid Nowak \altaffilmark{1}, Paul J. Wiita \altaffilmark{1},  Ann E. Wehrle \altaffilmark{2} and Stephen C.\ Unwin\altaffilmark{3}}
\email{revalsm1@tcnj.edu}
\altaffiltext{1}{Department of Physics, The College of New Jersey, PO Box 7718, Ewing, NJ 08628, USA}
\altaffiltext{2}{Space Science Institute, 4750 Walnut Street, Suite 205, Boulder, CO 80301}
\altaffiltext{3}{Jet Propulsion Laboratory, California Institute of Technology, Mail Stop 321-100, 4800 Oak Grove Drive, Pasadena, CA 91109}

\begin{abstract}
We have used photometry from the Kepler satellite to characterize the variability of four radio-loud active galactic nuclei (AGN) on timescales from years to minutes. The Kepler satellite produced nearly continuous high precision data sets which provided better temporal coverage than possible with ground based observations. We have now accumulated eleven quarters of data, eight of which were reported in our previous paper. In addition to constructing power spectral densities (PSDs) and characterizing the variability of the last three quarters, we have linked together the individual quarters using a multiplicative scaling process, providing data sets spanning $\sim 2.8$ years with $>98\%$ coverage at a 30 minute sampling rate. We compute PSDs on these connected data sets that yield power law slopes at low frequencies in the approximate range of $-1.5$ to $-2.0$, with white noise seen at higher frequencies.  These PSDs are similar to those of both the individual quarters and to those of ground based optical observations of other AGN.  We also have explored a PSD binning method intended to reduce a bias toward shallow slope fits by evenly distributing the points within the PSDs. This tends to steepen the computed PSD slopes, especially when the low frequencies are relatively poorly fit. We detected flares lasting several days in which the brightness increased by $\sim 15-20\%$ in one object, as well a smaller flare in another. Two AGN showed only small, $\sim 1-2\%$, fluctuations in brightness.
\end{abstract}

\subjectheadings{accretion disks --- galaxies: active --- galaxies: jets --- galaxies: photometry --- galaxies: Seyfert --- quasars: general}

\section{Introduction}
	Active Galactic Nuclei (AGN) are a class of galaxies characterized by intense emission from their central regions that exhibits variability on timescales from minutes to years in many wavebands. The standard physical model for the powerhouse of these objects involves material from an accretion disk falling onto a super-massive black hole (SMBH) (e.g., reviews by \citealt{antonucci93}, \citealt{urry95}). As the plasma infalls it is compressed and heated, emitting across much of the electromagnetic spectrum.  Some of the infalling material may be ejected at relativistic speeds via oppositely directed jets. These relativistic jets emit the synchrotron radiation which characterizes radio-loud AGN. Those AGN with jets pointed close to our line of sight often have their emission and variability significantly enhanced by relativistic Doppler boosting, producing rapid large scale variations and a flat radio spectrum dominated by the boosted core, leading to the blazar class of AGN (e.g., \citealt{scheuer79};  \citealt{kapahi82}). Observations of the variability of these AGN allow us to probe the inner portions of the accretion disk and/or jets close to the central SMBH engine (e.g., \citealt{mangalam93}; \citealt{marscher85}). As the variability of AGN should primarily originate in the accretion disk during quiescent periods, but be dominated by the relativistic jets during high activity states, the light curves they exhibit should be determined by the physical mechanism(s) causing the variability.\par

\begin{deluxetable*}{cccccccc}[H]
\tablenum{1}
\tabletypesize{\scriptsize}
\tablecaption{Kepler Active Galactic Nuclei Target List}
\tablehead{
\colhead{Object} & \colhead{Name} & \colhead{Kepler Input} & \colhead{Right} & \colhead{Declination} & \colhead{Kepler Input} & \colhead{Redshift} & \colhead{Radio} \\ 
\colhead{Designation} &  & \colhead{Catalog} & \colhead{Ascension} &  & \colhead{Catalog} & & \colhead{Spectral} \\ 
&  & \colhead{Number} & \colhead{(hh:mm:ss.s)}  &  \colhead{(dd:mm:ss)}  & \colhead{Magnitude} &  & \colhead{Index\tablenotemark{a}}}
\startdata
A & MG4 J192325+4754 & 10663134 & 19:23:27.24 & 47:54:17.0 & 18.6 & 1.520 & 0.32 \\
B & MG4 J190945+4833 & 11021406 & 19:09:46.51 & 48:34:31.9 & 18.0 & 0.513 & 0.75 \\
C & CGRaBS J1918+4937\tablenotemark{b} & 11606854 & 19:18:45.62  &  49:37:55.1 & 17.8 & 0.926 & 0.00 \\
D & [HB89] 1924+507 & 12208602 & 19:26:06.31  &  50:52:57.1 & 18.4 & 1.098 & 0.19 \
\enddata
\tablenotetext{a}{Radio spectral index from VLBA Calibrator website, defined between 2.3 and 8.3 GHz or 2.3 and 8.6 GHz with S $\propto \nu^{-\alpha}$.}
\tablenotetext{b}{Kepler Input Catalog incorrectly indicates that this target is a star with contamination 0.73, but we have verified it is an isolated quasar.}
\end{deluxetable*}

	The primary mission of the Kepler space telescope was to search for exoplanets around other stars, and to quantify how many may be in the star's habitable zone \citep{borucki10, koch10}. Kepler was capable of monitoring $\sim 150,000$ targets at one time, including nonstellar targets. The advantages of this space based observatory made Kepler an excellent platform from which AGN research could be carried out. Through the Kepler Guest Observer Program (Programs GO20018 and GO30010), we have obtained photometric data on four AGN  with an excellent sampling rate (30 minute and 1 minute cadences) and comparable precision to,  as well as a much higher duty cycle than, is commonly available employing ground based optical telescopes. Prior to the start of the Kepler mission, only approximately a dozen AGN were known in the Kepler field, including Cygnus A and II Zw 229-015. This scarcity of AGN was primarily due to the close proximity of the observed field to the galactic plane, which is not normally well covered in extragalactic surveys because of the galactic extinction and difficulty distinguishing galaxies from the copious foreground stars. Aside from our search for radio-selected AGN in the Kepler field, \citet{edelson12} devised a technique employing data from the WISE, 2MASS, and Rosat surveys that effectively finds Seyfert galaxies and brought the total number of AGN identified to $\sim 40$ with some three dozen more unconfirmed candidates. \par
	The Kepler satellite has been employed previously to determine the light curves and variability properties of AGN.   \citet{mushot11} reported on four Seyfert 1 galaxies, including the previously identified Seyfert, II Zw 229-015, for between two and four quarters.  All four of their Seyferts showed surprisingly steep red-noise power spectral density (PSD) slopes for each quarter that ranged from $-2.6$ to $-3.3$ at lower frequencies ($\sim 10^{-6.8}$Hz to $\sim 10^{-4.5}$Hz) and flattened out to white-noise spectra at higher frequencies extending up to $\sim 10^{-3.6}$Hz.  Further investigations of II Zw 229-015 were carried out by \citet{carini12} who made use of their overlapping ground based observations to determine a more optimal photometric aperture and connect four quarters of data.  This allowed for more accurate re-extraction of the Kepler photometric data, better distinguishing core variability by excluding more of the surrounding host galaxy.  They thus probed lower frequencies and found a somewhat shallower slope for the red-noise portion of the PSD.  Most recently, \citet{edelson13} described two quarters of Kepler data for the only highly variable BL Lacertae object that has been found in the Kepler field, W2R1926+42.  It showed large scale rapid flaring along with quiescent periods, and they found a substantial RMS-flux correlation. They also found that a broken power law fit to the PSD was not sufficient and suggested that more sophisticated mathematical models were needed to characterize this remarkable blazar. \par
	The primary purpose of this paper is to search for and characterize variability in the complete Kepler data sets spanning 11 quarters ($\sim 2.8$ years) for our four target AGN. The three newest quarters are also examined individually for variability. We connect, or stitch together, the 11 quarterly data sets we obtained for our four radio-loud AGN, three of which are quasars and the fourth presently identified as a Seyfert 1.5 galaxy. We also develop and apply new data analysis techniques including PSD binning which evenly populates the PSD points to remove a possible bias toward shallower slope fits, and light curve binning in order to estimate overall errors on our fitted PSD slopes.\par
	Our previous work involving the use of Kepler to study AGN included the processing of eight quarters of data on these four AGN \citep{wehrle13}, hereafter referred to as Paper I, will be summarized in the next section. That section also includes key data on our AGN targets. We then proceed to provide a description of our three most recent quarters of data in \S3 and show the PSDs for those quarters in \S4. This is followed by a discussion of the process by which we connect our quarterly data in \S5. Details of our binned PSDs are presented in \S6 and the consideration of short light curve segments designed to determine errors in the PSDs is in \S 7. We end with our results and conclusions in \S8.
\\
\section{Our Previous Work}
 Our targets within the Kepler field were selected based on the presence of core radio emission exceeding 100 mJy with a flat spectrum from the NRAO VLBA Calibrator List.   According to the unified model for AGN, these were expected to be flat spectrum radio quasars or BL Lac objects, with much of their optical flux originating from the Doppler boosted jet pointing rather close to our line-of-sight, and hence likely to be more variable on all timescales \citep{gopal03}. More details on the four targets so obtained and our selection criteria are discussed in Paper I. We display the key object information in Table 1. We refer the reader to Paper I for details of the following topics: target selection, detailed target information, Kepler satellite characteristics, instrumental effects, data artifacts, and our data correction process. Because the standard Kepler pipeline was designed to look for dips in light curves caused by exoplanets crossing stars of otherwise constant brightness, it removes too much of the inherent variability found in AGN.  Therefore we analyze the original SAP fluxes, but perform corrections that remove or reduce various instrumental effects. The various artifacts inherent in the Kepler data are detailed by the Kepler team \citep{keplerinstrument09}, as well as in \S5.3 of Paper I. We now briefly describe some key aspects of our observations as well as our data handling process.\par
	The Kepler satellite \citep{borucki10} is in an Earth trailing orbit, revolving around the Sun rather than the Earth.  Four times per year the satellite changed orientation in a quarterly roll in order to keep all solar panels directed at the Sun. These two factors caused a variety of instrumental effects. First, the telescope was constantly pointing in the direction of the constellation Cygnus. This means the pointing angle of the satellite was constantly changing with respect to Cygnus as it orbited the sun, changing the point spread function of the incoming light on the CCD camera.  This gave rise to differential velocity aberration at the few percent level. This effect manifests itself as an arc superimposed on the light curve over the course of a quarter. In addition, the Kepler satellite only stored data from predetermined apertures on the CCD camera to manage storage capacity. The ideal apertures for each target varied slightly from quarter to quarter, appearing as the overall jumps in electron counts between quarters as the number of pixels being summed changed after each quarterly reorientation. All four targets appear as point sources from ground based observations; however, the Kepler satellite deliberately slightly defocused images in order to obtain optimal photometry by spreading the light over several pixels to alleviate issues of pixel saturation and minor telescope motion.
	A few of the CCDs on which our targets fell were affected by unstable amplifier electronics which introduced a small nonlinear, quasi-periodic signal, known as the Moir{\'e} effect. This affected faint targets ($\sim$18th magnitude), such as ours, more severely, and the affected modules are listed in Table 13 of the \cite{keplerinstrument09}. See \S4.2 of Paper I for a further discussion of the effects on our targets, as well as \citet{Kolo10} for a technical discussion. Overall, less than 20\% of our data are affected by this phenomenon. \par

\begin{figure*}
\figurenum{1}
\centering
\subfigure
{\includegraphics[scale=0.27]{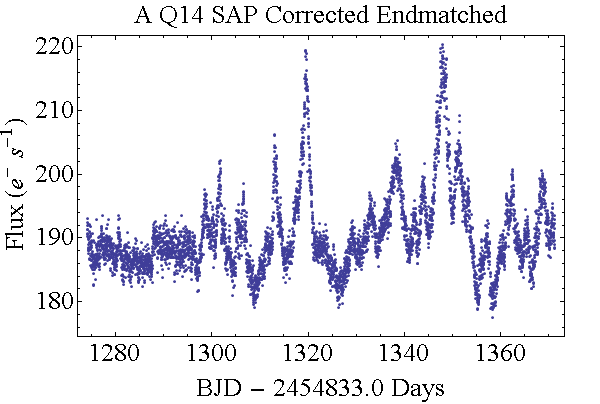}}
\quad
\subfigure
{\includegraphics[scale=0.27]{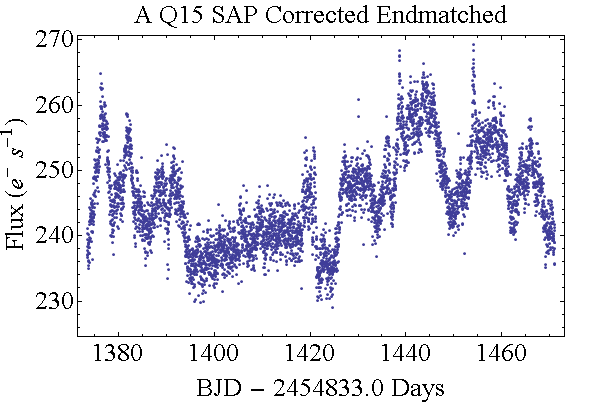}}
\quad
\subfigure
{\includegraphics[scale=0.27]{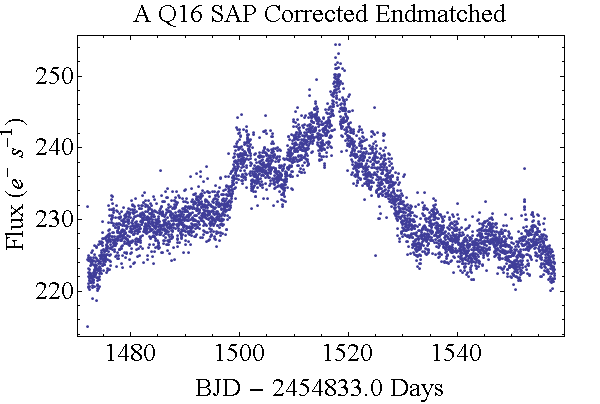}}
\quad
\subfigure
{\includegraphics[scale=0.27]{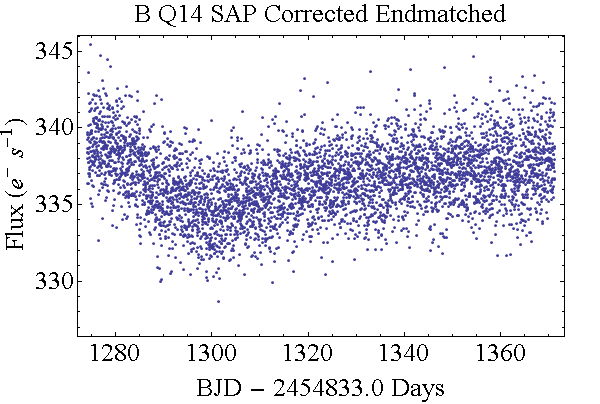}}
\quad
\subfigure
{\includegraphics[scale=0.27]{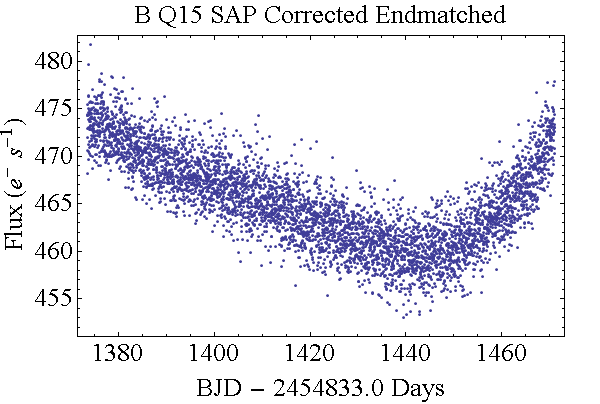}}
\quad
\subfigure
{\includegraphics[scale=0.27]{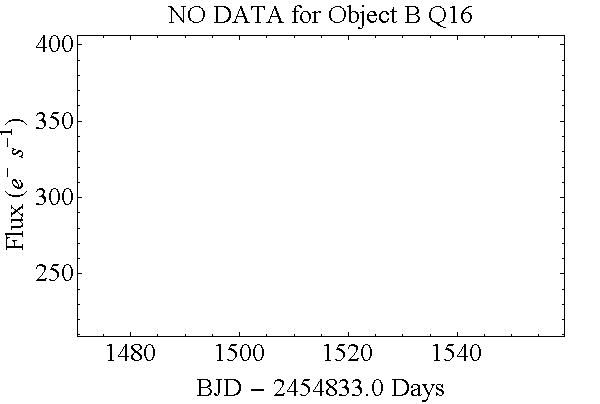}}
\quad
\subfigure
{\includegraphics[scale=0.27]{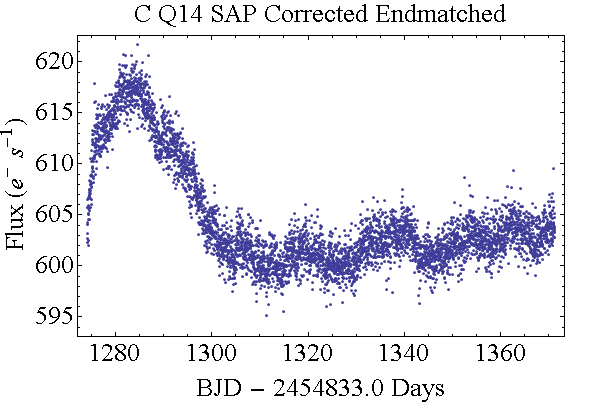}}
\quad
\subfigure
{\includegraphics[scale=0.27]{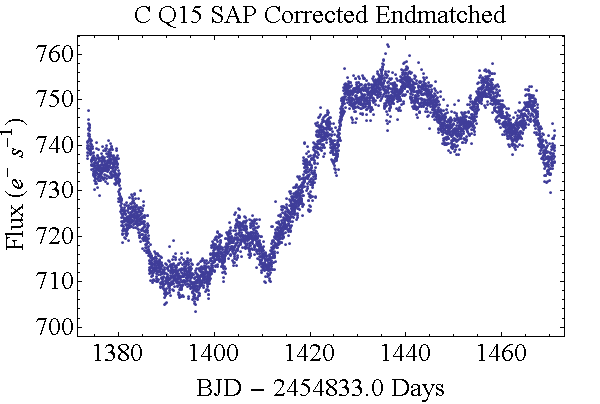}}
\quad
\subfigure
{\includegraphics[scale=0.27]{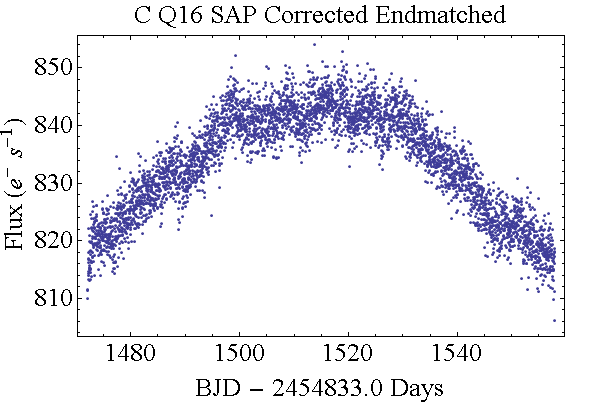}}
\quad
\subfigure
{\includegraphics[scale=0.27]{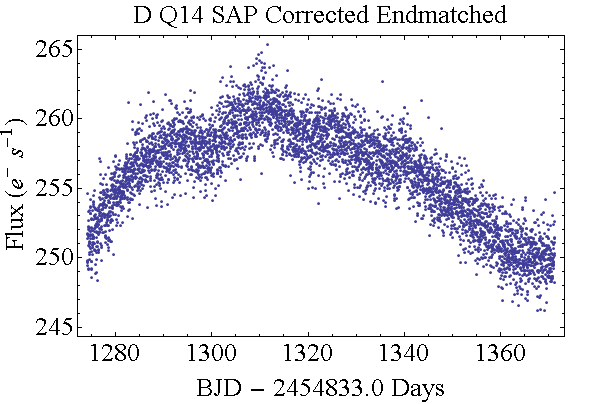}}
\quad
\subfigure
{\includegraphics[scale=0.27]{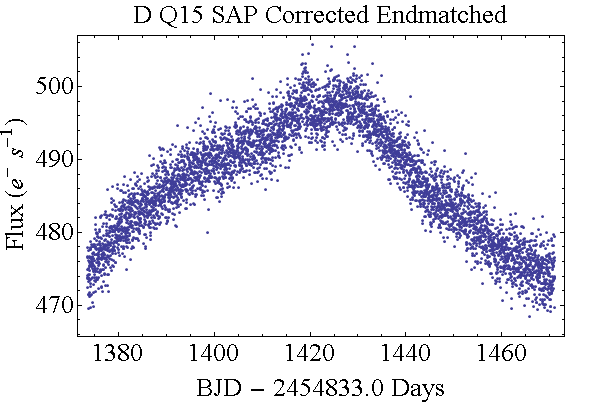}}
\quad
\subfigure
{\includegraphics[scale=0.27]{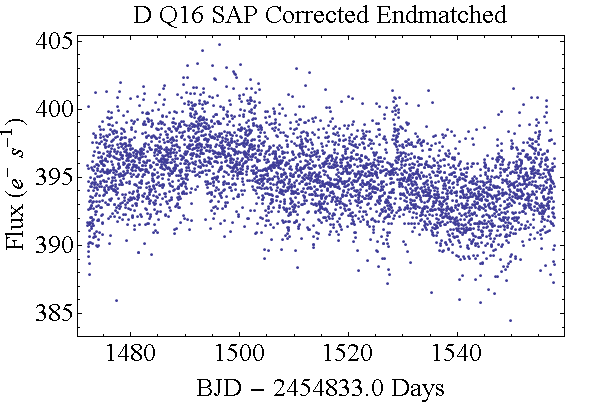}}
\quad
\caption{Light curves of the SAP corrected and endmatched data for Quarters 14 -- 16. There is no data for Object B during Q16. Considerable Moir{\'e} effects are noted in Object A during Q15, and very minor effects for A Q14, C Q15, and D Q14.}
\end{figure*}

\begin{deluxetable*}{ccccccccccccc}[H]
\tablenum{2}
\tabletypesize{\scriptsize}
\tablecaption{Power Spectral Density (PSD) Slopes for Quarter 14 - 16 Data}
\tablewidth{0pt}
\tablehead{
& & &\multicolumn{2}{c}{\textbf{Standard PSDs}} & & & \multicolumn{2}{c}{\textbf{Binned PSDs}}\\
& & \multicolumn{2}{c}{SAP Data} & \multicolumn{2}{c}{Corrected SAP Data} & \multicolumn{2}{c}{SAP Data} & \multicolumn{2}{c}{Corrected SAP Data} \\
\colhead{Object - } & \colhead{Moir{\'e}} & \colhead{Slope} & \colhead{Slope} & \colhead{Slope} & \colhead{Slope} & \colhead{Slope} & \colhead{Slope} & \colhead{Slope} & \colhead{Slope} \\
\colhead{Quarter} & \colhead{or RB} & \colhead{Original} & \colhead{End-Matched} & \colhead{Original} &  \colhead{End-Matched} & \colhead{Original} & \colhead{End-Matched} & \colhead{Original} &  \colhead{End-Matched}
}
\startdata 
A Q14 & Med & -1.9 & -1.7 & -1.9 & -1.8 & -1.5 & -1.4 & -1.5 & -1.4\\
A Q15 & High & -1.7 & -1.7 & -1.7 & -1.7 & -1.6 & -1.6 & -1.7 & -1.7\\
A Q16 & NR & -1.8 & -1.7 & -1.8 & -1.7 & -2.0 & -2.0 & -2.0 & -2.1\\
\tableline
B Q14 & NR & -1.6 & -1.0 & -1.8 & -1.1 & -1.6 & -1.6 & -1.7 & -2.1\\
B Q15 & NR & -1.6 & -1.5 & -1.6 & -1.6 & -1.8 & -1.8 & -1.9 & -2.0\\
B Q16 & *** & *** & *** & *** & *** & *** & *** & *** & ***\\
\tableline
C Q14 & NR & -1.9 & -1.6 & -2.0 & -1.9 & -1.7 & -1.8 & -2.0 & -2.6\\
C Q15 & Med & -1.8 & -1.5 & -2.0 & -2.0 & -1.8 & -1.8 & -2.1 & -2.4\\
C Q16 & NR & -1.9 & -1.6 & -1.9 & -1.5 & -1.9 & -1.9 & -1.9 & -1.9\\
\tableline
D Q14 & Med & -1.8 & -1.7 & -1.8 & -1.8 & -2.0 & -2.0 & -2.2 & -2.2\\
D Q15 & NR & -1.8 & -1.8 & -1.9 & -1.3 & -2.0 & -2.0 & -2.4 & -2.4\\
D Q16 & NR & -1.9 & -1.2 & -1.7 & -0.9 & -1.8 & -1.3 & -1.8 & -1.3
\enddata
\tablecomments{NR means no Moir{\'e} or Rolling Band effects reported; *** indicates no data was available.}
\end{deluxetable*}

	There are four remaining data artifacts which we have corrected for to obtain smooth light curves. First, we have written codes which fill in the rare individual null points within the data employing a line of best fit to the data and using the local standard deviation to fill in a reasonable point. Second, when the telescope turned to send back data each month, the CCDs suffered from a thermally induced focus shift manifesting as a discontinuity in the flux reading. These readings, however, decayed to pre-event levels as the telescope returned to normal operation. We fit and subtract out these glitches as exponential decays with time constants on the order of one to two days. Third, during the day it took to downlink all the data, no readings were being taken. We fill these two, one-day-long gaps each quarter using a similar method to that employed to replace the null points. Lastly, we remove the rare outlier points which are more than 4$\sigma$ away from the overall average and replace them with points determined by the local distribution. All analyses, including PSDs, are run on the data both with and without the above corrections made, but they do not make significant changes to the results (Paper I). \par
	We have observed all four targets in long cadence (30 minute sampling) mode for a total of 11 quarters, 8 of which were presented in our earlier paper.  We had short cadence data (1 minute sampling) available for two quarters for each object; however, as it provided little additional information for our targets and is expensive in terms of memory for the Kepler satellite to collect, no new short cadence data was obtained during the last three quarters.  We focus here on these three most recent quarters individually in order to characterize target variability as well as on linking all 11 quarters together for each AGN.

\section{Quarters 14 -- 16 Data Characteristics}
	We have analyzed newly acquired data on our AGN targets for Kepler Quarters 14, 15 and 16 (Fig.\ 1). Through examination of the light curves we note that during Q14 object A showed a large scale flare of $\sim 15\%$ from the baseline flux centered around Barycentric Kepler Julian Date (BKJD) 1320; BKJD is defined as BJD $-$ 2454833.0 days, and used so as to store smaller time stamps on the satellite. Throughout the quarter, multiple flares on the $\sim 5\%$ scale are noted. This flaring activity continued into Q15; however, object A unfortunately then fell on a Moir{\'e} affected CCD, preventing us from making any claims on the amplitude of the flaring activity then. During Q16, object A showed a few small variations, with two flares at the $\sim 6\%$ level. \par
	During Q14 object B was noted to show an $\sim 8\%$ dip in brightness consisting of four contiguous data points, centered on BKJD 1360. No loss of fine pointing or other instrumental glitch could be correlated with this peculiar event. Nonetheless, given the paucity of apparently intrinsic variations of this object, it is likely that this was an undetermined instrumental effect. During Q15 the primary observed variation for object B is due to velocity aberration. There is no data for object B during Q16 as that is one of the four quarters each year where it falls on dead CCD Module 3.\par
	During Q14 and Q15, object C showed continual variation at the $\sim 1-2\%$ level on a timescale of several days. During Q15 some of the variation may be due to the Moir{\'e} effect described in \S2. Throughout Q16 object C continued to show minor variations at the $\sim 1-2\%$ level.\par
	Object D was generally the least active, with only $\sim 1\%$ variations visible throughout the three quarters, superimposed on the large changes due to velocity aberration.\par
	Notable data artifacts include a Safe Mode event during Q14 for six days beginning on 14 July 2012. During this time no data was collected. A small solar coronal mass ejection is also noted on 25  June 2012 for $\sim 16$ hours but the effects were deemed insignificant by the Kepler team \citep{data19}. During Q15 one Safe Mode event occurred preventing data collection for $\sim 20$ hours. See the \citet{data20} for additional details.  In the case of both Safe Mode events we filled the gaps with points determined by a line of best fit through the data, imparting to the interpolated data a reasonable distribution based on the pre- and post-event standard deviations. As shown in Paper I, this makes little difference to the PSDs.\par
	Due to increasing friction on one of the Kepler reaction guidance wheels, the decision was made to have an 11.3 day rest period at the start of Q16. See the \citet{data22} for more information.
The permanent failure of this reaction wheel on 4 May 2013 meant that the data expected for Q17 could not be obtained. 

\section{Power Spectral Densities}
	We computed PSDs for our simple aperture photometry (SAP) flux both with and without our corrections applied.  In Paper I our PSD calculation was done using the program Period04 \citep{lenz05}; however, we now employ original codes which reproduce the same results much more efficiently. The use of our newly written codes and their agreement with our previous methods is reassurance that our methodology is not introducing systematics into the PSDs.\par
	We first fit the PSDs using the procedure detailed in Paper I. This consists of a least squares method in which we compute power-law fits starting from the lowest frequencies and advancing toward the higher frequencies until the quality of the fit significantly degrades, which indicates the break point between the low-frequency red noise and the high-frequency white noise. The break point is determined statistically by calculating the R-squared value of the fit line as each additional point is added to the fit. The differences in the R-squared values are stored, and the fit is terminated at the point before that difference is maximal. Given that the total number of points populating the PSDs is $\sim 50,000$ this process is relatively resistant to discontinuing the fit too early due to small spurious features in the PSD. We note that our slopes for all objects are consistent with that of our earlier 8 quarters of data, and the results, using this original method, are summarized in the first set of columns in Table 2. Overall, object B tends to have the flattest PSD slopes, with all other objects having low-frequency slopes consistently between $-1.6$ and $-2.0$.

	Over the past few years the Damped Random Walk (DRW) model has been employed to analyze quasar variability, typically for irregularly sampled light curves of large numbers of objects \citep{kelly09, kozlowski10, macleod10, zu13}. This method models the light curve as a stochastic random walk process with a damping term which pushes the light curve back toward its mean value (see \citet{ivezic14} for a concise review). The DRW model requires only three free parameters: the mean flux of the light curve, a characteristic timescale, and the asymptotic variability amplitude.  Using the DRW approach \citet{kelly09} found a strong correlation between the characteristic timescale and corresponding SMBH mass. \citet{kozlowski10} further developed this method as a selection tool for separating quasar light curves in large surveys. These results were confirmed by \citet{macleod10} using a larger sample size.  Most recently, \citet{zu13} explored additional parameters for more accurate modeling of quasar light curves.   The DRW model has the advantages of not suffering from uneven sampling and windowing biases that afflict standard spectral analysis methods.  However, thanks to the excellent quality of the Kepler data sets, combined with our corrections process, most of these problems are not an issue here.  Although implementing this model is beyond the scope of this paper, such an effort should be worthwhile. The basic SAP light curves are now publicly available from the MAST archive and our corrected light curves are available for further analysis to any interested parties by request to M.R. or P.J.W.

\section{Connecting Quarterly Data}
	In order to study AGN variability on longer timescales, a means must be found to connect the quarterly data together into one contiguous data set. The issues that arise in this process come from the manner in which Kepler collected data. Every quarter the satellite rolled $\sim 90$ degrees in order to maximize exposure of the solar panels to the Sun. Due to this, each object fell on four different CCD modules during the course of a year. Each module had its own predetermined optimal aperture for SAP.  This means the overall amount of flux being counted was slightly different from season to season, causing a baseline jump, or step, between adjacent quarters.  \citet{carini12} were able to use coordinated ground based observations to determine appropriate scaling of their quarters for a single source. For the majority of targets, including our four, this is not an option; however, we now have explored a scaling process intended to maintain the overall percentage of variations, while mitigating the issue of quarterly steps in baseline flux. \citet{edelson13} have used a similar approach in order to connect two quarters of data for their highly active BL Lac object.  We expanded this approach up to a temporal extent of 11 quarters. We continued to  employ end-matching, i.e., subtracting from the flux within each quarter a linear function of time so that the  flux values at the beginning and end of each  are equated, thereby removing the lowest order instrumental drift \citep{mushot11} and also providing a superior estimate of the power spectrum \citep{foug85}.

\subsection{``Stitching'' Algorithm}
	To accomplish the combination of quarterly SAP flux data we wrote two independent stitching codes to help reveal any broad systematic errors being introduced by the process. The procedure is as follows. First, we determine an overall average flux for all 11 quarters of data. Then we find the ratio between the mean of the first quarter of data and that average. This allows us to identify a multiplicative scaling factor that we use to scale the first quarter either up or down to the overall average. Following this original normalization we scale the remaining quarters to the preceding quarter by scale factors that we determine from the ratio of an average of the last twenty points in the first quarter of each pair to the first twenty points in the second quarter of the pair.  This process is designed to normalize each quarter to the mean for the entire data set, being a scaling process that maintains the true percentage of observed variations.\par
	We first combined the corrected SAP data for all 11 quarters without performing any additional corrections.  We then repeated this procedure for all 11 quarters of data for our targets with two separate end-matching corrections applied.  We combined quarters that had been individually end-matched and also ones that were not end-matched, followed by an overall end-matching of the combined 11 quarter light curve. The original and scaled data obtained using the first end-matching approach for our targets are shown in Fig.\ 2. These processes are intended to remove instrumental drift in varying amounts \citep{foug85}, as well as uncover any significant discrepancies in the methods used to determine the PSDs. We show the results in Table 3 and discuss the effects of the end-matching later.  Two independently written codes were used to conduct the above stitching and PSDs were calculated from both. The overall broken power-law slopes for the PSDs from each method almost always were consistent to within $\sim 0.03$.

\begin{figure*}
\figurenum{2}
\centering
\subfigure
{\includegraphics[scale=0.4]{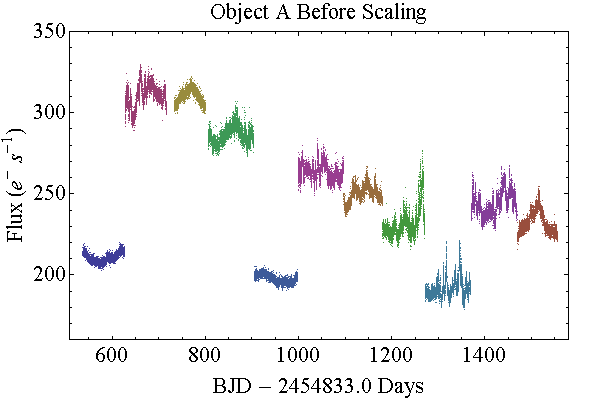}}
\quad
\subfigure
{\includegraphics[scale=0.4]{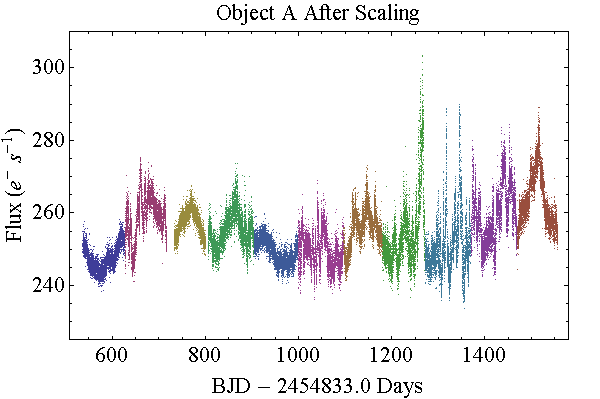}}
\quad
\subfigure
{\includegraphics[scale=0.4]{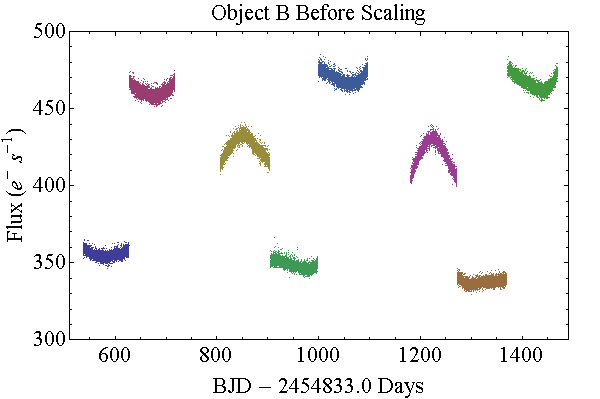}}
\quad
\subfigure
{\includegraphics[scale=0.4]{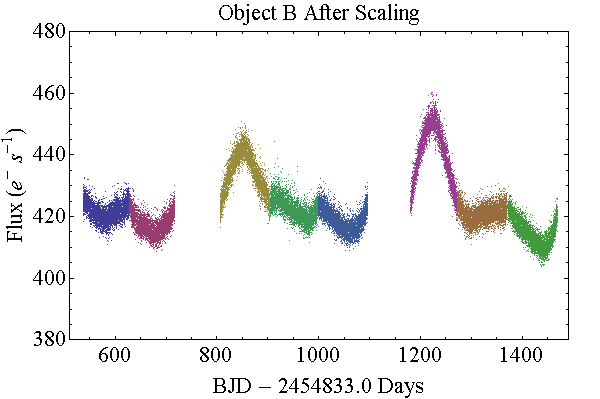}}
\quad
\subfigure
{\includegraphics[scale=0.4]{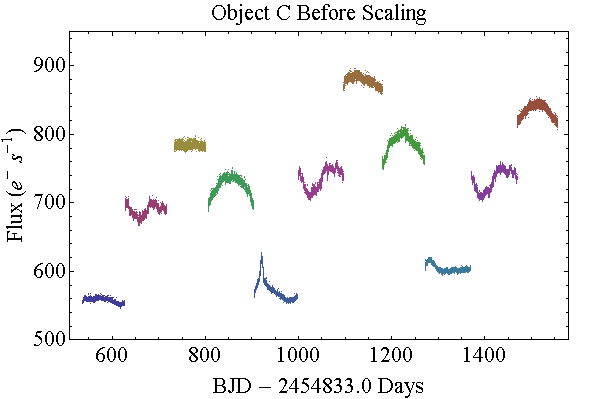}}
\quad
\subfigure
{\includegraphics[scale=0.4]{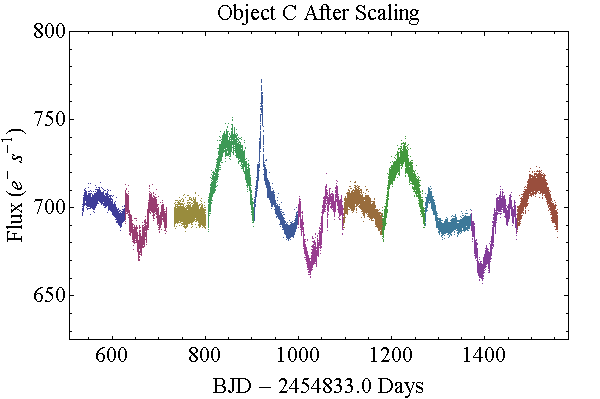}}
\quad
\subfigure
{\includegraphics[scale=0.4]{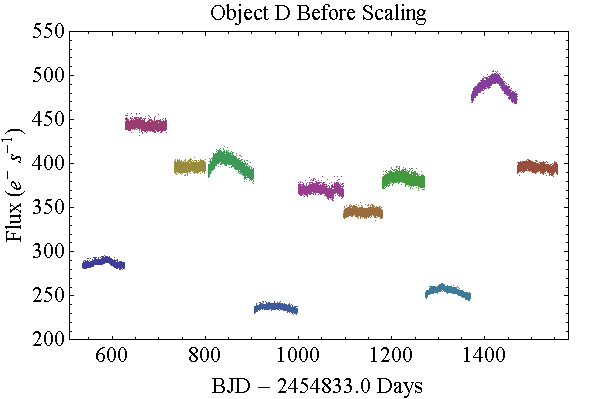}}
\quad
\subfigure
{\includegraphics[scale=0.4]{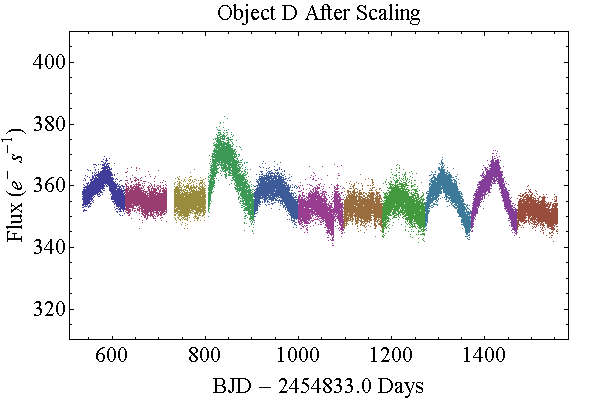}}
\quad
\caption{Combined Quarters 6 -- 16 SAP light curves of all four objects. Left: each quarter has been individually end-matched before scaling. Right: the final result of the scaling and connecting process. Our corrections have been applied for monthly thermal glitches and downlink gaps as well as rare null points. The axes are electron counts per second and Truncated Barycentric Julian Date.}
\end{figure*}

\begin{figure*}
\figurenum{3}
\centering
\subfigure
{\includegraphics[scale=0.4]{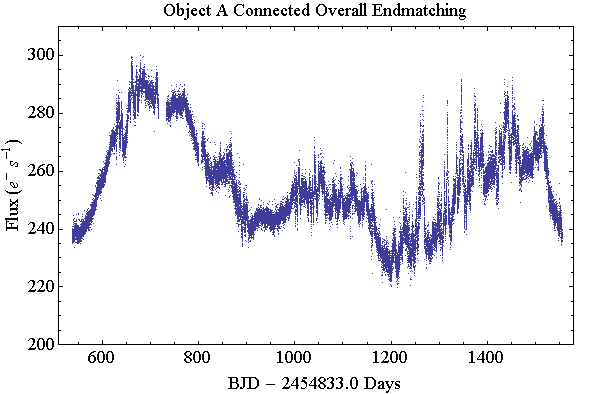}}
\quad
\break
\subfigure
{\includegraphics[scale=0.4]{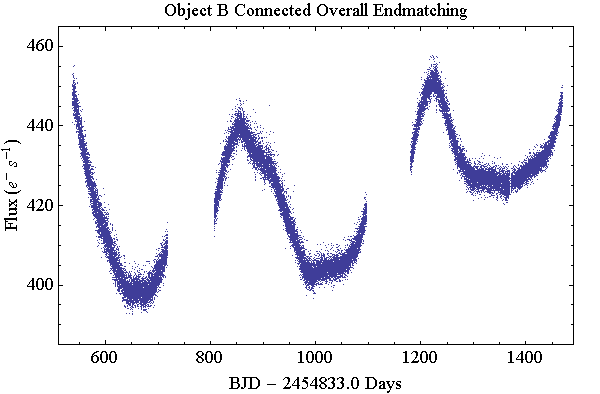}}
\quad
\break
\subfigure
{\includegraphics[scale=0.4]{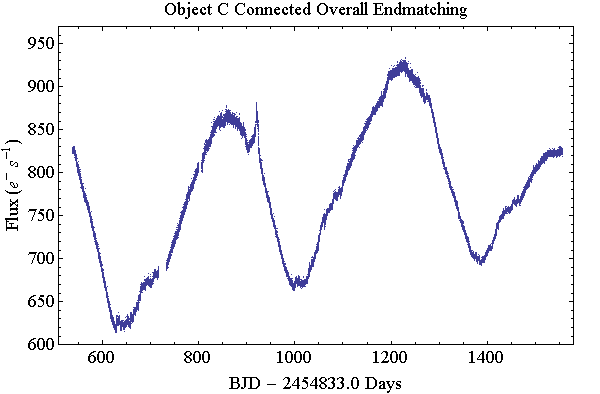}}
\quad
\break
\subfigure
{\includegraphics[scale=0.4]{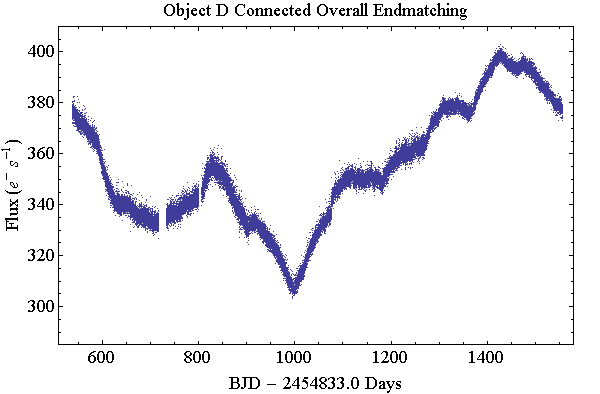}}
\quad
\caption{Combined Quarters 6 -- 16 SAP light curves of objects A-D. We have corrected for monthly thermal glitches and downlink gaps as well as rare null points. The overall light curves have been end-matched. The axes are electron counts per second and Truncated Barycentric Julian Date. The periodic trends strongly present in objects B and C are predominantly produced by differential velocity aberration.}
\end{figure*}

\begin{deluxetable*}{cccccccccc}
\tablenum{3}
\tabletypesize{\scriptsize}
\tablecaption{Power Spectral Density (PSD) Slopes for Q6 -- 16 Connected Data}
\tablewidth{0pt}
\tablehead{
& \multicolumn{2}{c}{Original PSDs} & \multicolumn{2}{c}{Binned PSDs} &\multicolumn{4}{c}{Quarterly PSD Averages \& Standard Devitations} \\
\colhead{Object} & \colhead{Slope} & \colhead{Slope} & \colhead{Slope} & \colhead{Slope} & \colhead{Slope} & \colhead{Slope} & \colhead{Slope} & \colhead{Slope}\\ 
\colhead{Name} & \colhead{Original} & \colhead{End-Match} & \colhead{Original} &  \colhead{End-Match} & \colhead{Original} & \colhead{SD} & \colhead{End-Match} & \colhead{SD}
}
\startdata 
A & -1.7 & -1.7 & -1.7 & -1.6 & -1.9 & 0.20 & -1.7 & 0.25 \\
B & -1.9 & -1.6 & -1.8 & -1.9 & -1.8 & 0.17 & -1.4 & 0.30 \\
C & -1.9 & -1.8 & -1.9 & -1.8 & -1.9 & 0.18 & -1.9 & 0.22 \\
D & -1.8 & -1.6 & -1.8 & -1.7 & -1.7 & 0.40 & -1.2 & 0.32
\enddata
\tablecomments{PSD Slopes for each object with all available 11 quarters of data connected using the process described in \S5.1. Here end-matched refers to overall end-matching of the final light curve, which is made up of individual quarters that were not end-matched. The quarterly averages were computed using the corrected non-end-matched and end-matched data sets.}
\end{deluxetable*}

\subsection{Connected Light Curves}
	The results of stitching together the 11 individual light curves for each of our objects yield $\sim 2.8$ yearlong data sets which exhibit a number of intrinsic as well as instrumental signatures.  These are presented in Fig.\ 3.  The overall light curves have been end-matched, but are composed of quarters that individually were not end-matched.  In all four objects, particularly objects B and C, there is a periodic effect due to differential velocity aberration superimposed upon the true variation of each object. Overall, objects A and C show the most true variations, with object A the most active.\par
	Object A shows moderate amplitude variations at the $\sim 5-20\%$ level throughout the $\sim 2.8$ years of collected data. While some quarters exhibit variations due to the Moir{\'e} effect (see \S2), unaffected quarters clearly show moderate and rapid changes in baseline flux. Overall this object has been steadily increasing over time in both overall activity, and amplitude of the noted flares.
	Object B generally shows the least variation of all four objects, with the only clear variation due to differential velocity aberration. There are two major gaps in the data during Quarters 8 and 12, and an early end to the data train (no Quarter 16), since object B fell on the dead CCD Module 3 at those times.
	Object C shows small scale variations at the few percent level with a high duty cycle. There is one major flare lasting for several days noted around KBJD 925. See Fig.\ 2 of Paper I for more details.
	Object D also displays variations primarily attributed to differential velocity aberration with sporadic variations at the few percent level.\par
Each light curve still contains the approximately day long gaps between quarters, comprising $\sim 1\%$ of the data. Overall, the averaging of points at the beginning and end of adjacent quarters appears to appropriately match up each quarter with the following one, with no major disjoints noted. We note the process is not ideal in a small number of cases, such as between Quarters 13 and 14 for object A, where the former ended on the downward trend of the largest flare, so perhaps in this case more points might have been preferable for the averaging.  Nonetheless, this multiplicative scaling appears to  work very well overall, as seen in the light curves of our objects as well as the highly active BL Lac investigated by \citet{edelson13}.

\section{Power Spectral Density Binning}
	All points in the PSDs computed in Paper I and in \S 4 above are weighted equally; however, the distribution of points is not equal due to the logarithmic nature of the plot, and there are approximately ten times as many points in any given decade of frequency as compared with the previous lower decade of frequency. This can introduce a bias into the fitting process and evenly distributing the points used to compute the PSD will alleviate this possibility.  Such a binning before computing PSDs has been used previously by \citet{carini12} as their primary method for computing PSDs.
	Our original fitting method is successful despite this bias due to the rapid change in the quality of the fit once the break point to instrumental white-noise is reached. Thus, our overall fitting tends to fit all frequencies below the break equally well and we see no evidence for curvature in the PSDs.  Nonetheless, in the cases where fits are not ideal, the general trend is for the lowest frequencies to be more poorly fit, as the larger number of points found toward the higher frequencies will dominate and might cause the fitted slopes to be shallower than the actual slopes.\par
	To examine the possible extent of this bias, we have binned our PSD values into equally spaced logarithmic bins before recomputing the slopes; both non-binned and binned PSDs for the full end-matched light curves for our targets are shown in Fig.\ 4. This means there are an increasing number of points per bin moving toward higher frequencies. We then average each bin, and replace all points in each bin with a single averaged point. The result of this procedure is that we then have evenly spaced data points populating our PSD, and this should remove any bias toward shallower fits. The slight disadvantage of this approach is that we now have only $\sim 4$ points per decade in frequency, so there is a reduction in our resolution of the determination of where the break point between red-noise and white-noise falls. This loss of break frequency precision appears to be minimal and overall the fits to the binned PSDs are nearly always as good, if not better, than those found without such binning.  The resulting slopes tend to match, to a high degree of accuracy, the slopes fitted to the standard PSDs when the low frequencies are well fit. In the cases were the lowest frequencies are not well fit, the binning tends to steepen the slope slightly, as anticipated, and fits the low frequencies and overall red noise better.
	The resulting binned PSD slopes are summarized in the last four columns of Table 2, in which binned PSD slopes are given for both the original data and the processed (``corrected'') data. They can be compared to the non-binned PSD slopes given in the preceding four columns. The power-law fits to the red-noise portion of the binned data are all very good and indicate that any apparent rollovers to flatter slopes at the lowest available frequencies that appear to be present in the unbinned PSDs are not real for our sources, and can be attributed to small number statistics.

\begin{figure*}
\figurenum{4}
\centering
\subfigure
{\includegraphics[scale=0.37]{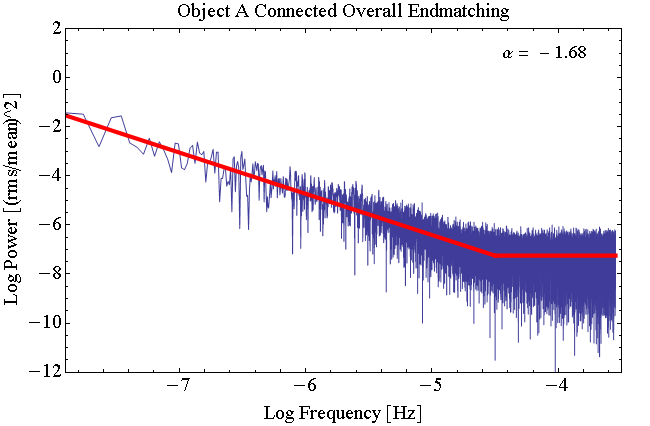}}
\quad
\subfigure
{\includegraphics[scale=0.37]{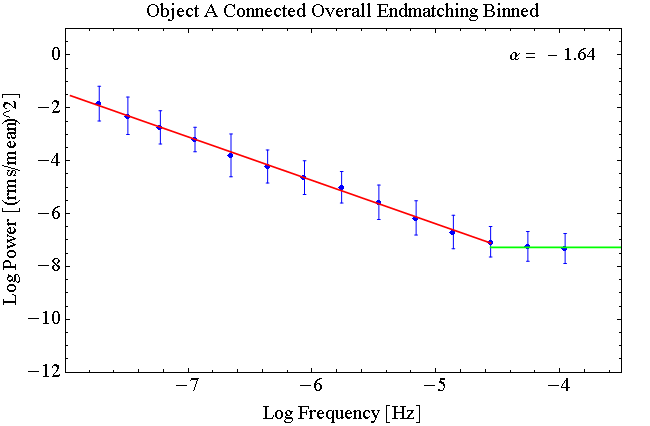}}
\quad
\break
\subfigure
{\includegraphics[scale=0.37]{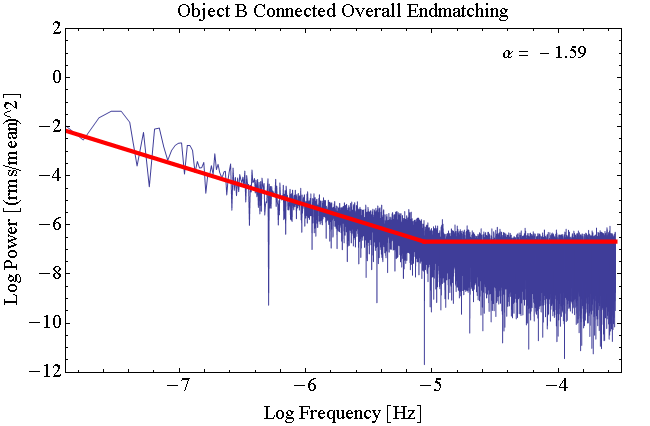}}
\quad
\subfigure
{\includegraphics[scale=0.37]{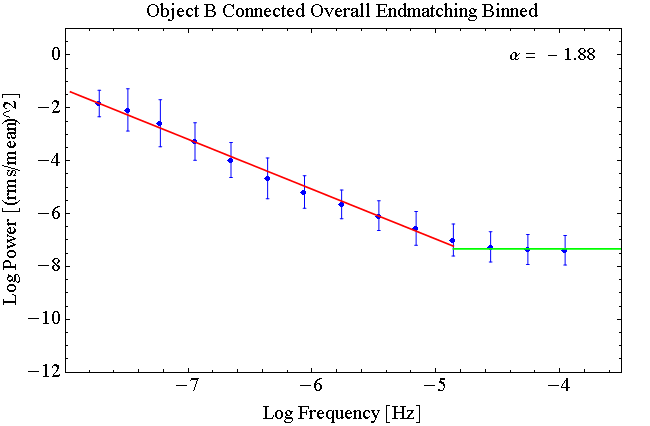}}
\quad
\break
\subfigure
{\includegraphics[scale=0.37]{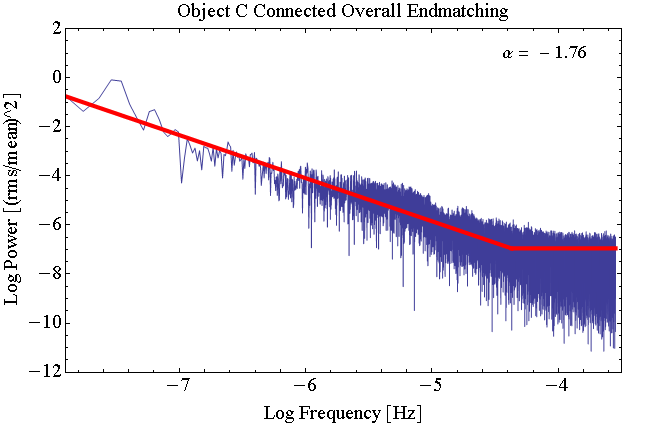}}
\quad
\subfigure
{\includegraphics[scale=0.37]{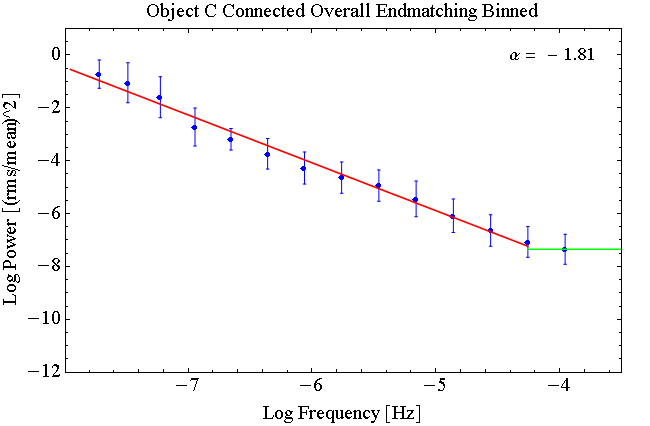}}
\quad
\break
\subfigure
{\includegraphics[scale=0.37]{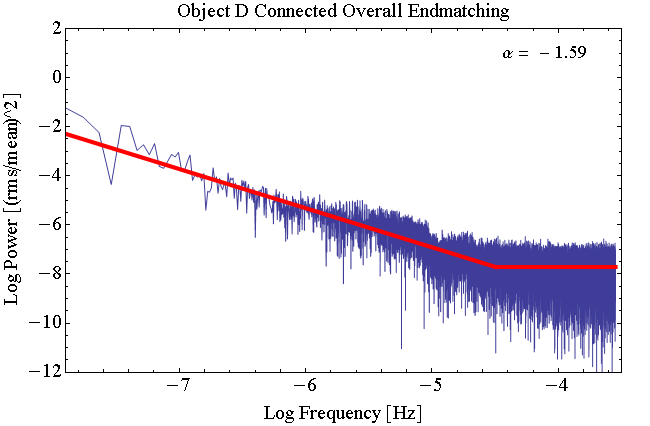}}
\quad
\subfigure
{\includegraphics[scale=0.37]{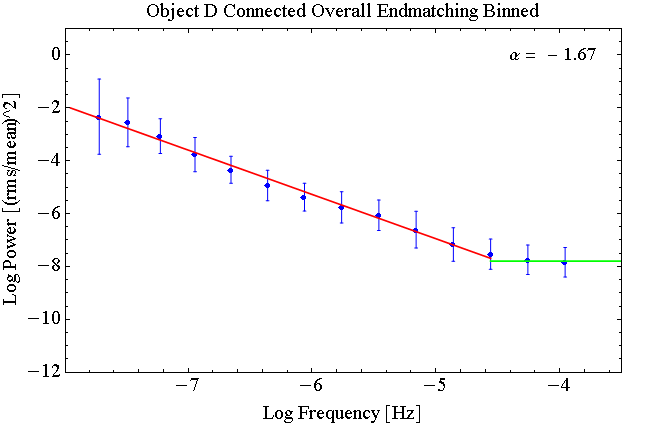}}
\quad
\caption{Standard and binned PSDs of the connected Kepler light curves. The slopes ($\alpha$)  of the low-frequency portions of the binned PSDs are shown.}
\end{figure*}

\section{Light Curve Sections}
	After computing average PSD slopes for their complete set of Kepler observations, \citet{edelson13} also divided their light curves into $\sim 10$ day sections, for which PSDs were calculated individually.  The principle idea behind this procedure is to see if the physical processes at work would reproduce similar PSD slopes on timescales shorter than the overall data set despite being calculated over a shorter range of temporal frequencies. Assuming the physical processes producing the variations are not changing and the short segments are fair samples of the overall dataset, variations in the PSDs from bin to bin allow overall errors to be quoted on the PSD slopes.
	To examine this point we broke our overall stitched together light curves into ten day segments, excluding any with quarterly gaps. This produced  $\sim 80$ ``bins'' for each of three of our AGN; the exception is object B, for which there are no data for one quarter per year due to the dead CCD Module 3, reducing the approximate number of bins to 60.\par
	Overall we found quite large variations between the PSD slopes computed from these limited frequencies, and shallower slopes on the whole. While the slopes for objects A and C were shallower, although comparable to the quarterly slopes, objects B and D were significantly shallower on average.  These results are summarized in Table 4.  We attribute the differences to removing too much low frequency power when the data are broken up into 10 day segments for our AGN, which are considerably less active than others such as the BL Lac discussed by \citet{edelson13}. In addition, for our targets such brief temporal segments do not probe sufficiently into the low frequency portion of their PSDs, resulting in shallower slopes.\par
	When comparing the slopes of the 11 individual quarters' PSDs to the overall stitched data there is excellent agreement. As we have 11 quarters of data for each object, we have a large enough sample size over which to average and still obtain a reasonable error estimate.  The last sets of columns in Tables 3 and 4 illustrate these points.

\begin{deluxetable}{cccc}
\tablenum{4}
\tablecaption{Ten Day Light Curve Bin PSD Averages}
\tablewidth{0pt}
\tablehead{
\colhead{Object} & \colhead{End Correction} & \colhead{Mean Slope} & \colhead{Standard Deviation}
}
\startdata 
A & Not End-Matched & -1.24 & 0.51 \\
A & End-Matched & -1.31 & 0.57\\
\tableline
B & Not End-Matched & -0.41 & 0.37 \\
B & End-Matched & -0.18 & 0.40 \\
\tableline
C & Not End-Matched & -1.40 & 0.34 \\
C & End-Matched & -0.98 & 0.57 \\
\tableline
D & Not End-Matched & -0.48 & 0.40 \\
D & End-Matched & -0.22 & 0.36
\enddata
\tablecomments{Averages computed from Q6-15 data sections without major breaks such as quarterly rolls. Each section was excised from the final stitched together light curves. See \S7 for a discussion. The quarterly slope averages yield a much better agreement with the slopes of the stitched together data.}
\end{deluxetable}

\section{Discussion and Conclusions}
 	We find the PSD slopes calculated for the connected light curves to be very consistent with the slopes of the individual quarters themselves. Thus it appears likely that our codes are appropriately linking the quarters together. Twenty points, each corresponding to a 30-minute integration cadence, are averaged over in determining the values used to connect the ends of each quarter; in some cases this is more than sufficient, but occasionally more or fewer points may be appropriate. We may explore a statistical optimizing method to determine the ideal number of points to average over in the future.\par
	Not surprisingly, the two AGN whose light curves show little variability on ten-day timescales have shallower PSD slopes when their individual ten-day intervals are fitted, as compared to those AGN with considerable variability. Objects A and C, which tend to be the most active, have 10-day interval PSD slopes which more closely agree with the overall quarterly slopes as compared with our other two objects. It is quite likely that the dominant observed timescales for our objects are significantly longer than for the BL Lac measured by \citet{edelson13} either because of the fundamental timescale set by the SMBH mass or the Doppler factor of the relativistic jet.\par
	We typically find quarterly and stitched PSD slopes between $-1.6$ and $-2.0$ for our AGN, with excellent fits based on a broken power law. We do not find any evidence of a knee or bending power laws in the red-noise portion of the PSDs for our targets as seems to have been found for the highly active BL Lac described by \citet{edelson13}. This may be due to the overall low activity of our objects, supported by the overall higher break frequencies noted for their object. Upon our own investigation of the BL Lac object, we found a broken power law was an excellent fit to the quarter 14 PSD, but inadequate for quarters 11 and 15. This supports that a bending power law model or the DRW model may be more appropriate for some AGN.\par
	On the whole, end-matching tends to yield shallower slopes for the standard quarterly PSDs, by $\sim 0.2$ on average. For the binned quarterly PSDs, the overall effect of end-matching is negligible. We note a similar trend for the connected light curve PSDs, with the standard PSD slopes being shallower by $\sim 0.15$ following end-matching. Again, the effects of the end-matching are negligible for the binned PSDs.\par
	We find binning the PSDs to evenly distribute data alleviates occasional poor line fits and apparently shallower slopes at the lowest frequencies we probed. Overall, binned slopes tend to be the same as those originally found, except in cases where the break point was not very clear and the lowest frequencies were poorly fit. This tends to steepen the slopes on a case by case basis but only by $\sim 0.12$ on average.\par
	In conclusion, we have analyzed three additional quarters of Kepler photometric data on four radio-loud AGN. We also combined all 11 quarters of data together in a manner which scales all variations to a single average and preserves the fractional size of variations. Through this process we have obtained light curves that are $\sim 2.8$ years long with a sampling rate of 30 minutes and a remarkable $>$ 98\% duty cycle. The validity of these results is supported by noting the overall PSD slopes are consistent with those of the individual quarterly data. The methodology employed here is expandable to any number of quarters and is only limited by the quantity of data available.\par
	 At higher frequencies the PSDs are dominated by instrumental white-noise. The computed low-frequency red-noise PSD slopes are similar to those of models predicting variability from turbulence behind a shock in relativistic jets \citep{marscher91} or accretion disks \citep{mangalam93}.  However, these models have not yet been developed to the point where even the extremely long and high quality data sets provided by Kepler allow us to discriminate between them. Once such models have been improved and can be contrasted with such excellent data sets, a clearer picture of the physical origin of AGN variability should emerge.

\section*{Acknowledgements}
We thank Victoria Calafut for assistance in developing original PSD codes and scripts. We are grateful to Paolo Di Lorenzo, Daniel Silano, and Daniel Sprague for developing codes and techniques used in Paper I that were also employed here. We thank Brandon Kelly for providing his DRW code and the anonymous referee for comments which improved the scope of this manuscript. We acknowledge support from the NASA Kepler Guest Observer Program through Grants NNX11B90G and NNX12AC83G (AEW, PI) and from the 2013 Mentored Undergraduate Summer Experience (MUSE) Program at TCNJ (PJW, PI) during which most of this work was carried out. Part of this research was carried out at the Jet Propulsion Laboratory, California Institute of Technology, under a contract with the National Aeronautics and Space Administration.

\end{document}